# FITS Foreign File Encapsulation Convention

Nelson Zarate (NOAO)
Rob Seaman (NOAO)
Doug Tody (NRAO)

10 September  2007

## 1  Introduction

This document describes a FITS convention developed by the IRAF Group (D. Tody, R. Seaman, and N. Zarate) at the National Optical Astronomical Observatory (NOAO).  This convention is implemented by the fgread/fgwrite tasks in the IRAF fitsutil package.  It was first used in May 1999 to encapsulate preview PNG-format graphics files into FITS files in the NOAO High Performance Pipeline System.

## 2  FOREIGN File Extension

A FITS extension of type 'FOREIGN' (henceforth a "FOREIGN file extension" or just "FOREIGN extension") provides a mechanism for storing an arbitrary file or tree of files in FITS, allowing it to be restored to disk at a later time.  Each FOREIGN extension contains a single file.   This mechanism also provides a means for associating a group of FITS extensions of any type.  Certain of the file attribute keywords can be included in the header of any FITS file or extension to support such things as storing a directory tree containing images, tables, and other non-FITS types of files in a multi-extension FITS file, and later restoring the whole tree to disk.  The motivation for this extension is to allow an implementation based on the FITS multi-extension mechanism to encapsulate and pass non-FITS data.   The FOREIGN extension may be used to store a file from any type of operating system (e.g. UNIX or Windows), however some of the specific file attributes that are recorded in the FOREIGN extension keywords may not map completely between different systems (e.g. the UNIX filemode string that may be recorded in the **FG_FMODE**  keyword does not have an exact counterpart under Windows).

The header of a FOREIGN FITS extension must begin with the following five keywords in the specified order with no intervening keywords.

```
  1   XTENSION= 'FOREIGN '
  2   BITPIX  =                    8
  3   NAXIS   =                    0
  4   PCOUNT  =           <filesize>  / file size in bytes
  5   GCOUNT  =                    1
      .
      EXTNAME = '<filename>'
```

Some early implementations of the FOREIGN extension reversed the order  of the PCOUNT and GCOUNT keywords, but this usage is now deprecated.  The optional EXTNAME keyword is used only to identify the extension in listings.  To restore a file to disk the "FG" (file group) keywords are used as outlined in the following section.

# 3  File Group (FG) Keywords

To be able to later unpack FOREIGN extensions and restore files to disk, a number of keywords must be added to the extension headers to store the information required to restore the files.  These are the "FG" keywords.  The FG keywords are used in both FOREIGN type extensions and in standard FITS extensions such as IMAGE, BINTABLE, and so on.

> **FG_GROUP** (string) - Each time a file group is written a group name is assigned. The group name associates all of the elements of a group.  Assuming the group name is unique then this can be used to associate all the extensions in a group for later restoration.  This is useful if groups are concatenated in a larger sequence of extensions.  The group name is arbitrary (like a filename) and is assigned by the user when the file group is written.  For example, a group name for a directory tree might be the name of the root directory.  It is up to the writer program to assign a group name if the user does not predefine one.
>
> **FG_FNAME** (string) - The filename of the file associated with the current extension. The maximum filename length is 67 characters. Any printable character except apostrophe is permitted.  For an extension of type FOREIGN where the file type is directory, FG_FNAME is the name of the directory.
>
> **FG_FTYPE** (string) - The physical file type.  The following types are recognized:
> - "text" - A file containing only text.  Stored 8 bits per character using newline to delimit lines of text (like Unix).
> - "binary" - Any file which is not a text file or one of the known file types.  Stored as a byte stream without any conversion.
> - "directory" - implementation dependent
> - "symlink" - implementation dependent
> - "FITS" -  a native FITS extension
> - "FITS-MEF" - a native multi-extension FITS (MEF) file.  No count of the number of extensions in the MEF file is given, rather the MEF group consists of all subsequent extensions until a FITS extension is encountered which starts a new file.
>
> **FG_MTYPE** (string) - The logical or "mime" type of the file (optional).
>
> **FG_LEVEL** (integer) - The directory nesting level.  All of the files in a directory are at the same level.  FOREIGN extensions of type directory are used to name the directories at each level so that pathnames can be reconstructed (this scheme assumes that the extensions in a file group are ordered).  Level 0 (zero) is the root directory of the file group.  The root directory is unnamed and is implicitly the user's current working directory into which the file in the FOREIGN extension would be unpacked.   When packing files into FOREIGN FITS extensions, the current working directory could be a logical choice for the **FG_GROUP** file group name.
>
> **FG_FSIZE** (integer) - The size in bytes of the data portion of the file.  This value is always identical to the value of the **PCOUNT** keyword.  In the case of a file  with a **FG_TYPE** value equal to "directory", the **FG_FSIZE** value is zero.
>
> **FG_FMODE** (string) - The file mode as a string ("rwx-rwx-rwx", bits not set given as "-").

**FG_FUOWN** (string) - The file UID (user ID) as the file owner name string.

**FG_FUGRP** (string) - The file GID (group ID) as the file group name string.

**FG_CTIME** (string) - The file creation time as a UTC value expressed as an ISO 8601 string.

**FG_MTIME** (string) - The file modification time as a UTC value expressed as an ISO 8601 string.

**FG_COMP** (string) - This keyword will not be used initially, but is reserved in case we choose to implement file (e.g. gzip) compression in the archiver. The value would be a string such as "none" or "gzip". In the meantime files can be archived in compressed form by compressing them beforehand and archiving the compressed files as binary files. Part of the reason we are reluctant to implement compression in the archiver is that archive data may last indefinitely and it is hard to guarantee that the compressed data will be readable a decade or two in the future. We might need to avoid compression for archival data unless the compression algorithms and/or code are part of the archive as well. (This discussion refers only to foreign files, not to compressed images).

# 3  Examples

The following examples are taken from actual runs on the IRAF implementation of this convention using the tasks *fgread* and *fgwrite* from the external package *fitsutil.* These utilities are written in C and are not tied to any IRAF system library.

The IRAF *fitsutil* external package contains the fgwrite/fgread tasks to write and read FOREIGN extensions. These are scripts that call the native C programs fgwrite.e and fgread.e with the following arguments:

```
fgwrite [-t <tbdsfm>] [-o <tbdsfm>] [-vdih] [-g <group_name>]
            [-f output_fits_file] [input_files]
```

Switches:
```
    f     write to named file, otherwise write to stdout
    d     print debug messages
    v     verbose; print full description of each file
    g     FG_GROUP name.  The default is the root directory name
    t     select file types to include in the output file
    o     skip file types from input files selection
    h     do not produce primary HDU
    i     write Table Of Contents in primary HDU
    s     calculate CHECKSUM and DATASUM for the input file
```

```
fgread [-t <tbdsfm> [-o <tbdsfm>] [-n ranges] [-vdxrf]
             [-f fitsfile] [files]
```

where ranges is of the form 1,2,5,8-11

Switches:
```
    d     print debug messages
    f     read from named file rather than stdin
    n     get list of extension numbers to extract
    o     omit the indicated FITS types (tbdsfm)
```

```
              r       replace existing file at extraction
              s       check CHECKSUM if keywords are present
              t       include the indicated FITS types (tbdsfm) only
              v       verbose; print full description of each file
              x       extract files

              The possible file types are
                      t: text
                      b: binary
                      d: directory
                      s: symbolic link
                      f: single FITS  file
                      m: multiple extension FITS file (MEF)
```

Example 1:   Create a FITS file containing an arbitrary set of files in a directory.

```
fi> dir r* l+
 -b-rwrwr- zarate      3616 Aug 14  9:23 rdf.o
 -t-rwrwr- zarate      6489 Aug 14 10:30 rdf_plio.c
 xt-rwrwr- zarate      6952 Aug 14 10:31 read_plio
 xt-rwrwr- zarate      6903 Aug 13 14:43 read_plio_save
 xt-rwrwr- zarate      6952 Aug  8 15:02 readf_save

fi> fgwrite r* /tmp/fg.fits   # Create a FITS file with the above files

fi> fxh /tmp/fg.fits  # See the FITS file contents (single line output)
EXT#    EXTTYPE             EXTNAME         EXTVER      BITPIX
0       /tmp/ft.fits                                    8
1       FOREIGN             rdf.o           1           8
2       FOREIGN             rdf_plio.c      1           8
3       FOREIGN             readf_save      1           8
4       FOREIGN             read_plio       1           8
5       FOREIGN             read_plio_save 1           8
```

Example 2:   Here are the values of some of the FG_ keywords for the case where the FITS file contains files that were originally in the tki/ directory and the tki/dir2 subdirectory.

```
EXT#   FG_FNAME     FG_FTYPE    FG_LEVEL FG_FSIZE   FG_FMODE      FG_FUGRP

0
1      tki          directory   1        0          drwxrwxr-x    users
2      max.o        binary      2        1616       -rw-rw-r--    users
3      dir2         directory   2        0          drwxrwxr-x    users
4      list.txt     text        3        69         -rw-rw-r--    users
5      home.txt     text        3        1113       -rw-rw-r--    users
6      gmttolst.c   text        2        1243       -rw-r--r--    users
7      a.c          text        2        770        -rw-rw-r--    users
8      varg.c       text        2        284        -rw-rw-r--    users
9      max.c        text        2        372        -rw-rw-r--    users
```

# 4 Implementation Notes

The following design notes refer to the fgwrite and fgread tasks in the IRAF fitsutil package, and provide some additional context and background information relating to the original motivations for the FOREIGN extension.

The fgwrite and fgread programs as used in the telescope data handling system are host callable (Unix) level tasks.

Sample syntax:

> fgwrite <flags> <input-file-template-list>
>
> fgread  <flags> <input-file>

The intention is not to provide a general file archive capability, but rather to be able to use FITS to carry along and archive some non-FITS auxiliary data. A secondary goal is to generalize FITS somewhat so that directories can be handled (archived and later restored) as well as linear file templates.

Since the goal is not to provide a general file archive capability, certain details are not addressed:  symlinks to directories are not followed by the writer;  unlike tar, hard links are not preserved;  special files are ignored.

Selected task options:

- Input-file-template-list is a sequence of file names or directory names (if it is a unix task, any templates will already have been expanded by the shell).
- There should be an option to fgwrite specify the types of files to be archived; when descending a directory, a file list alone will not handle this.  Hence some mechanism such as which of the possible supported file types (tbdsf), or a pattern matching template such as in "find -name", would be used to select the files to be archived.

Output File Format

The output host file (or byte stream) is a conventional FITS file consisting of a sequence of one or more FITS extensions, optionally preceded by a dataless  primary header unit (PHU) describing the entire file.  Writing of the PHU may be disabled even if a file is being written to disk (e.g. when writing a sequence of extensions to be concatenated).

Foreign files (text, binary, directory, symlink) are wrapped as single extensions with XTENSION='FOREIGN'. Single FITS images without extensions are converted to IMAGE extensions, writing a single extension to the output stream.

Multi-extension FITS files in the input are written unchanged except that keywords are added to the first HDU to identify the MEF group (subsequent extensions are merely copied to the output stream unchanged). If the first HDU in the input file is a PHU it is converted to an IMAGE extension.  The order of the extensions in the output stream must match that in the input MEF for the MEF to be later restored to disk. The PHU and all extensions in the input MEF are still visible in the output file; their association as an MEF grouping is evident only by examining the FG keywords in the HDU.  Any internal MEF associations, such as for inheritance, are still present, but might not be recognized by most software until the MEF group is later restored to a file.

By default the output stream will have a dataless PHU describing the contents of the file (this can be disabled as mentioned above).  The PHU may optionally include a table of contents for the output file. If a TOC is generated this will require that the output file list be fully processed to determine the type and size of each input file, before writing out the PHU with TOC followed by the input data files. This might be desirable in any case to simplify the code (construction of the input file list can be separated from file conversion and output).